\documentclass{aa}  

\usepackage{graphicx}
\usepackage{txfonts}
\usepackage{natbib}
\usepackage{amsmath}
\usepackage[pdfauthor={},pdftitle={},colorlinks=true]{hyperref}
\usepackage{array}

\usepackage{xcolor}
\usepackage{ulem}

\newcommand{\hms}[3]{\ensuremath{{#1}^\mathrm{h}{#2}^\mathrm{m}{#3}^\mathrm{s}}}
\newcommand{\dms}[3]{\ensuremath{{#1}^\mathrm{\circ}{#2}\mathrm{'}{#3}\mathrm{''}}}
\newcommand{\heim}{{\tt heimdall}}
\newcommand{\frb}{FRB\,121102}

\begin{document} 


   \title{Constraints on the low frequency spectrum of \frb}

   \subtitle{}

   \author{L.~J.~M. Houben\inst{1}\fnmsep\inst{2}\fnmsep\thanks{\email{l.houben@astro.ru.nl}}
        \and L.~G. Spitler\inst{2}
        \and S. ter Veen\inst{3}
        \and J.~P. Rachen\inst{1}
        \and H. Falcke\inst{1}\fnmsep\inst{3}\fnmsep\inst{2}
        \and M. Kramer\inst{2}\fnmsep\inst{4}}

   \institute{Department of Astrophysics/IMAPP, Radboud University, PO Box 9010,
              6500\,GL Nijmegen, The Netherlands
        \and
              Max-Planck-Institut f\"ur Radioastronomie, Auf dem H\"ugel 69,
              D-53121 Bonn, Germany
        \and
              ASTRON, the Netherlands Institute for Radio Astronomy, Oude Hoogevensedijk 4,
              7991 PD Dwingeloo, The Netherlands
        \and
              Jodrell Bank Centre for Astrophysics, University of Manchester,
              Alan Turing Building, Oxford Road, Manchester M13 9PL, United Kingdom}

   \date{}

   \abstract{While repeating fast radio bursts (FRBs) remain scarce in number, they provide a unique opportunity for follow-up observations that enhance our knowledge of their sources and potentially of the FRB population as a whole. Attaining more burst spectra could lead to a better understanding of the origin of these bright, millisecond-duration radio pulses. We therefore performed $\sim$\,20\,hr of simultaneous observations on \frb\ with the Effelsberg 100-m radio telescope and the Low Frequency Array (LOFAR) to constrain the spectral behaviour of bursts from \frb\ at 1.4\,GHz and 150\,MHz. This campaign resulted in the detection of nine new bursts at 1.4\,GHz but no simultaneous detections with LOFAR. Assuming that the ratio of the fluence at two frequencies scales as a power law, we placed a lower limit of $\alpha > -1.2 \pm 0.4$ on the spectral index for the fluence of the instantaneous broad band emission of \frb. For the derivation of this limit, a realistic fluence detection threshold for LOFAR was determined empirically assuming a burst would be scattered as predicted by the NE2001 model. A significant variation was observed in the burst repeat rate $R$ at L-band. During observations in September 2016, nine bursts were detected, giving $R = 1.1 \pm 0.4$\,hr$^{-1}$, while in November no bursts were detected, yielding $R < 0.3$\,hr$^{-1}$ (95\% confidence limit). This variation is consistent with earlier seen episodic emission of \frb. In a blind and targeted search, no bursts were found with LOFAR at 150\,MHz, resulting in a repeat rate limit of $R < 0.16$\,hr$^{-1}$ (95\% confidence limit). Burst repeat rate ratios of \frb\ at 3, 2, 1.4, and 0.15\,GHz are consistent within the uncertainties with a flattening of its spectrum below 1\,GHz.
}

   \keywords{Methods: data analysis - Stars: neutron - Stars: activity}

   \maketitle

\section{Introduction}
In the past few years, several bright, millisecond-duration, and highly dispersed single pulses have been observed with radio telescopes across the world~\citep[e.g.][]{pbj+16}. This new class of radio transients, called fast radio bursts (FRBs), have dispersion measures (DMs) greater than a few times the DMs expected from our Galaxy~\citep{cl03}, placing their yet unknown sources outside the Milky Way and presumably at cosmological distances. To explain the distances and energetics involved, a wide variety of progenitor models have been proposed, many of them associating FRBs to cataclysmic events~\citep{tot13,kim13,fr14,gh15}. 

Such models cannot explain FRBs like \frb\ that are seen to repeat~\citep{ssh+16-1,ssh+16-2}. Although it might not be representative of the full class of FRBs, it demonstrates the value of a precisely known position. After the detection of additional bursts consistent in location and DM with the initial discovery burst of \frb~\citep{sch+14}, a direct sub-arcsecond localisation of a repeat burst was achieved using the Karl G. Jansky Very Large Array (VLA)~\citep{clw+17}. This allowed \citet{tbc+17} to identify the host galaxy of the source and to confirm its extra-galactic origin. Optical and infrared observations further showed the host to be a faint, low-metallicity dwarf galaxy at a redshift of $z = 0.193$~\citep{tbc+17} and the source to be located in a bright star-forming region~\citep{bta+17}. The VLA data also revealed a persistent radio source at the position of \frb~\citep{clw+17}. European Very Long Baseline Interferometry Network (EVN) observations subsequently showed that bursts of \frb\ originate within 40\,pc of this source, which suggest the two are physically connected~\citep{mph+17}. The bursts are also 100\% linearly polarised after a Faraday rotation correction with a rotation measure (RM) on the order of RM $\textasciitilde 10^{5}$ rad\,m$^{-2}$~\citep{msh+18}. Such a rotation measure, which was also seen to vary, requires a highly magnetic and ionised environment.

Since the properties of this persistent radio source resemble those of a low-luminosity, accreting massive black hole~\citep{mph+17}, \frb's bursts could originate from a source in the vicinity of a massive black hole. The source could be a neutron star that then produces short-duration repeated emission like the giant pulses of radio pulsars~\citep{cw16,lbp16}. Alternatively, the bursts could originate from a young magnetar~\citep{lyu02,lyu14} and the persistent radio source is explained as a pulsar wind nebula~\citep{km17} or a supernova remnant~\citep{mbm17}.

No periodicity has been found within the detected radio bursts~\citep{ssh+16-2,hds+17,zgf+18}, which is in contrast to what is expected for bursts originating from a rotating neutron star. The bursts of \frb\ are reported to be correlated in time~\citep{ssh+16-2,lab+17,oyp18} where the detection of one burst implies a higher probability of detecting another shortly thereafter. Its burst spectra also seem to be band-limited with a full with half maximum (FWHM) of approximately 500\,MHz and a central frequency that varies from burst to burst~\citep{ssh+16-1,ssh+16-2,lab+17}. It is as yet unclear if these properties are intrinsic to the source or caused by propagation effects~\citep{cwh+17}.

In order to further constrain burst spectra, this work presents efforts to detect \frb\ simultaneously with the Effelsberg 100-m radio telescope, hereafter denoted with ``Effelsberg'', and the Low Frequency Array (LOFAR). Section \ref{sec:observations} describes the observations. A detailed description is given on how the resulting data were searched for single pulses in Sect. \ref{sec:analysis}. The results of this single pulse analysis are presented in Sect. \ref{sec:results}, which are then used for a multi-frequency burst comparison including previously published bursts in Sect. \ref{sec:burst-comp}. We conclude this work with Sect. \ref{sec:conclusion}.

\section{Observations}\label{sec:observations}
Between 4 September and 14 November 2016, four simultaneous observations were performed with Effelsberg and LOFAR of roughly five hours each. Within the five hour blocks, a few pulsars were observed for the purpose of calibration and to test the functionality of our triggering system. This triggering system, further described in Sect. \ref{sec:obsLOF}, was based on a private \mbox{VOEvent} Network~\citep{ad09} connecting the two observatories and enabled a rapid response to a potential Effelsberg detection with LOFAR.

Table \ref{tab:obsprop} gives an overview of the \frb\ observations, their overlap, and how many bursts have been found in each of them. During the September observations, both telescopes were centred at the sky position RA = \hms{05}{31}{58}, DEC = \dms{33}{08}{04} (J2000)~\citep{ssh+16-1}. In November, Effelsberg was pointed at the updated sky position of \frb~\citep{mph+17} (RA = \hms{05}{31}{59}, DEC = \dms{33}{08}{50}), while LOFAR was centred at the previously used position. Assuming the primary beams of both telescopes to be Gaussian in shape, this positional error introduced a drop in sensitivity relative to a bore-sight detection of $\sim$2\% for a FWHM Effelsberg beam of 0.166$^{\circ}$ and $\sim$7\% for a FWHM LOFAR core tied-array beam of $\sim$5$'$. These values are smaller than the overall uncertainty on the sensitivities of the receivers and therefore neglected in all forthcoming calculations.

\begin{table}
\caption{Details about the performed observations.}
\label{tab:obsprop}
\centering
\begin{tabular}{l l >{\centering\arraybackslash}p{1.5cm} c >{\centering\arraybackslash}p{0.8cm}}
   \hline\hline
   \noalign{\smallskip}
   Tel. & Date & Start time (UTC) & Duration & No. bursts\\
   \noalign{\smallskip}
   \hline
   \noalign{\smallskip}
   EFF & 2016-09-04 & 04:51:30 & 4h 24m & 5\\
   LOF & 2016-09-04 & 04:30:00 & 5h 00m & 0\smallskip\\
   EFF & 2016-09-19 & 04:37:43 & 3h 52m & 4\\
   LOF & 2016-09-19 & 04:30:00 & 3h 56m & 0\smallskip\\
   EFF & 2016-11-13 & 01:24:59 & 4h 35m & 0\\
   LOF & 2016-11-13 & 02:27:00 & 4h 42m & 0\smallskip\\
   EFF & 2016-11-13 & 23:28:37 & 6h 31m & 0\\
   LOF & 2016-11-13 & 23:12:00 & 4h 42m & 0\\
   \noalign{\smallskip}
   \hline
\end{tabular}
\tablefoot{EFF = Effelsberg, LOF = LOFAR.}
\end{table}

\subsection{Effelsberg}\label{sec:obsEFF}
The Pulsar Fast Fourier Transform Spectrometer (PFFTS) was used to record high time resolution, total intensity spectral data (i.e. Stokes-I only), similar to that for the Northern High Time Resolution Universe (HTRU North) pulsar survey~\citep{bck+13}. This search backend yielded data with a total of 512 frequency channels corresponding to a bandwidth of 300\,MHz centred at 1360\,MHz and a time resolution of 54.613\,$\mu$s. It recorded data for all the beams of the 7-beam feed array so as to be able to better discriminate between radio frequency interference (RFI) and actual bursts from \frb.

Data from the central beam was also recorded using the high-precision pulsar timing backend PSRIX~\citep{lkg+16}. Although the sensitivity of the PFFTS spectrometers are higher, they are not synchronised with a precise clock and do not provide accurate timestamps. The PSRIX data, with a bandwidth of 250\,MHz centred at 1358.9\,MHz divided into 256 channels and a time resolution of 51.2\,$\mu$s, were therefore used to determine the exact times of arrival (TOAs) of the bursts detected with Effelsberg.

~~~~~

During the observations, the Effelsberg data were analysed in real time to allow the timely submission of FRB \mbox{VOEvents}~\citep{phb+17} to LOFAR. For this purpose the raw 32-bit PFFTS data from the central beam of the receiver were processed in gulps of 16 seconds with the graphics processing unit (GPU) -based transient detection pipeline \heim\footnote{\url{https://sourceforge.net/projects/heimdall-astro/}.}~\citep{thes:b12}. It was set to search for single pulses between a DM of 0 and 2000\,pc\,cm$^{-3}$. DM steps ranging from 0.2 to 11\,pc\,cm$^{-3}$ were used to convolve the dedispersed time series with a set of boxcars with widths $2^n \times t_{samp}$, where $0 \leq n \leq 9$ and $t_{samp}$ is the sampling time of the data to search up to a width of 27.96\,ms. Candidates were selected with a signal to noise ratio (S/N) of at least 6 and visually inspected for bursts at DM = 560\,pc\,cm$^{-3}$. For every candidate deemed significant by the observer, a trigger was sent to LOFAR initiating the transient buffer board (TBB) response, described in the next sub-section, within $\sim$30\,s after the detection of the burst.

\subsection{LOFAR}\label{sec:obsLOF}
The high-band antennae (HBAs) of LOFAR~\citep{hwg+13} were used to store visibility data and time series in the frequency range of 110 to 188\,MHz. For reasons explained below, the 14 remote stations and one core station were set to the interferometric imaging mode that produces correlated raw visibility data, and the remaining 23 core stations were set to the beamforming mode. This configuration produced one tied-array beam with an angular size of $\sim$5$'$, which is the coherent sum of all the core station beams centred at the position of \frb. The standard pulsar search pipeline~\citep{sha+11} was used to convert the 8-bit beamformed data into PSRFITS formatted time series with a time resolution of 1.31\,ms. To reduce the effect of intra-channel dispersion smearing, 25600 frequency channels were used to cover the 78\,MHz of bandwidth. Bursts of \frb\ would be smeared to $\sim$2, $\sim$4, and $\sim$11\,ms at 190, 150, and 110\,MHz, respectively.

Additionally, the raw complex antenna voltages of the HBAs were continuously recorded on the TBBs implemented at all core and remote stations of LOFAR. The TBB ring buffers can store 2\,GB of data for each dipole, preserving a 5.2-second history of the radio sky. In order to use these data, the TBB recording must be stopped (``frozen'') by an internal or external trigger and then, after some potential further analysis of the trigger, read out (``dumped'') to the central LOFAR computer for offline processing. Due to network restrictions and the large amount of data to be read out, a dump can generally only be performed once per observation. Afterwards TBB data can be coherently dedispersed to yield resolved pulse profiles and imaged to localise the origin of bursts down to several arcseconds using the longest of the remote station's baselines~\citep{vef+18}. This makes TBB data valuable for FRB research, but as the dump of a full TBB buffer takes significant time during which the system is off, freezing or dumping TBB data must be decided on rigid trigger criteria.

During our observations, the TBBs' viability for FRB observations was tested by providing triggers from Effelsberg via a prototype implementation of the \mbox{VOEvent} standard for FRBs described by \citet{phb+17}. Within fractions of a second upon the reception of a trigger at LOFAR, the TBBs were automatically frozen at a time most likely to detect the dispersed signal with the TBBs. For \frb\ this is possible as the high DM of the source causes its pulses to arrive at 188\,MHz about one minute after they are recorded at 1.4\,GHz. The decision whether to dump frozen TBB data onto disk was then made after a more detailed analysis of the corresponding Effelsberg candidate to ensure it was real. This validation was only possible for the brightest Effelsberg bursts, hence a total of two dumps were performed as a response to bursts V and IX in Table \ref{tab:burstprop}.

To analyse single pulses within the TBB data, an accurate calibration of the time structure of the signal is required. \citet{cbe+16} studied this in detail for sub-microsecond pulses produced by cosmic rays for which the LOFAR TBBs are primarily used~\citep{snb+13}, but how to calibrate TBB data for the detection of FRBs is less well understood. Imaging our raw visibilities, recorded with the remote stations and a single core station, using the standard LOFAR techniques~\citep{hbh+11} could yield time delay tables that can be applied one-to-one on the TBB data. This procedure would significantly simplify the TBB data calibration, but requires a signal of \frb\ to be found in the TBBs. As the sensitivity of the TBBs is roughly the same as for the beamformed data, finding a pulse in the latter is a precondition for further analysis of the TBB data. It is therefore the analysis of the recorded beamformed data that is described below and which leads to the results presented in this paper.

\section{Analysis}\label{sec:analysis}
\subsection{Effelsberg data}
Complementary to the real-time search, the 32-bit PFFTS data from all the seven beams were converted offline into 8-bit filterbank data to conserve computation and storage resources and re-analysed with \heim. This time, pulses were searched over a DM range of 0 to 5000\,pc\,cm$^{-3}$ to allow the detection of a new FRB in any of the receiver's beams. The local RFI was identified by comparing candidates from the outer beams with those from the central beam, and removed from the candidates list. Remaining promising events were then confirmed to be real bursts from \frb\ by visual inspection of the raw filterbank data using the task {\tt waterfaller.py} from the software package PRESTO\footnote{\url{http://www.cv.nrao.edu/~sransom/presto/}}~\citep{thes:r01}.

To ensure no bursts were missed by \heim, the 8-bit filterbank data were also analysed using PRESTO alone. After making an RFI mask with the PRESTO task {\tt rfifind}, the data were dedispersed with DM steps of 1\,pc\,cm$^{-3}$ from 530 to 590\,pc\,cm$^{-3}$, a range centred around the DM of 560\,pc\,cm$^{-3}$ that most optimally aligns the sub-structure of the bursts of \frb\ and is therefore considered to be its true DM~\citep{hss+18}. The data were downsampled by a factor of 16 to attain a time resolution comparable to the data's inter-channel smearing of $\sim$1\,ms. Similar to \heim, the time series were convolved with several boxcar match-filters and candidates were created for events with an S/N $>$ 6 using the PRESTO task {\tt single\_pulse\_search.py}~\citep{lbh+15}. The same nine bursts were found with both methods and are presented in Sect. \ref{sec:results}.

The characteristics of these bursts, that is their radio peak flux densities and pulse widths, were determined by extracting a small snapshot of dedispersed time-frequency data around the bursts with the digital signal processing for pulsars (DSPSR) program~\citep{sb11}. The bandpass of the receiver was removed from the resulting PSRCHIVEs, the data products of DSPSR, before they were averaged over frequency and downsampled in time with a factor that yielded the best least squares fit to the bursts' profiles using a Gaussian model. The widths and heights of the bursts were determined from these fits. For each burst, its height was converted to a peak flux density by applying the radiometer equation for two summed polarisations, 235\,MHz of bandwidth, due to band roll-off and flagged channels, and its S/N and width determined by the fit~\citep[Eq. A1.21 in][]{book:lk12}. These peak flux densities, widths, and S/Ns are listed in Table \ref{tab:burstprop}.

We recall from Sect. \ref{sec:obsEFF} that the PFFTS data have no absolute time information. In order to improve the TOAs of the bursts in this data, determined as an offset in seconds from the start of their corresponding filterbank file, the same PRESTO analyses as applied to the 8-bit filterbank data was applied to the PSRIX data. This search yielded TOAs for seven bursts. For the two faintest bursts in the PFFTS data, no counterpart was found in the PSRIX data. Nevertheless a TOA was determined for all nine bursts by adding a correction obtained from bursts found in both data sets to the arrival times in the PFFTS data. The TOAs given in Table \ref{tab:burstprop} are the barycentred TOAs of the bursts referenced to infinite frequency calculated from the topocentric TOAs and updated sky position of \frb\ with the PRESTO function {\tt bary}. These arrival times enabled a targeted search for burst counterparts in the LOFAR band.

\subsection{LOFAR beamformed data}\label{sec:analysis-lof}
Although the persistent source surrounding could become optically thick at higher frequencies, free-free absorption is stated to be negligible for \frb\ at LOFAR frequencies by \citet{tbc+17}. \frb\ is affected by scattering and scintillation, the effects of which are seen to be consistent with the predictions of the NE2001 model~\citep{cl03} as caused by a Galactic electron column density of $\sim$188\,pc\,cm$^{-3}$~\citep{mph+17,msh+18}. LOFAR bursts from \frb\ are therefore expected to be strongly scatter-broadened to $\sim$35, $\sim$100, and $\sim$385\,ms at 190, 150, and 110\,MHz, respectively, assuming Kolmogorov frequency scaling ($\tau \propto \nu^{-4.4}$). In order to accommodate such pulse profile evolution over frequency in our search for LOFAR counterparts to the Effelsberg bursts, a new search method had to be devised for the LOFAR data.

Unlike the Effelsberg data, the beamformed data were searched in eight sub-bands of $\sim$10\,MHz, for scattering might otherwise wash out the highly dispersed pulses from \frb\ when the entire bandwidth is frequency-scrunched to a time series. Additionally, shorter bursts can be resolved in the higher frequency sub-bands due to less intra-channel smearing in these bands (see Sect. \ref{sec:obsLOF}). With the PRESTO task {\tt prepsubband,} eight sub-bands were created for 100 dispersion trials between a DM of 545 and 575\,pc\,cm$^{-3}$ suppressing present RFI using an RFI mask made with {\tt rfifind}. Because {\tt prepsubband} cannot downsample sub-bands individually, all sub-bands per dispersion trial were downsampled by a factor of two, attaining a time resolution of 2.62\,ms compared to a total minimum smearing of 3.89\,ms for the highest frequency sub-band. The dispersion trials per sub-band were then converted to time series using {\tt prepsubband} again, as if eight sub-bands had been dedispersed individually. Each sub-band was then searched for bursts with {\tt single\_pulse\_search.py} using wider boxcar widths the lower the frequency range of the sub-band up to a maximum width of 300 times the time resolution of the data (i.e. 786\,ms).

%
%

The resulting single pulse candidates with an S/N higher than 8 were visually inspected in a DM versus time diagram. In such a diagram, symmetric tear-drop shaped groups of candidates are a characteristic for broad band single pulses, as shown by \citet{cm03}, but none were evident. Many of the sub-bands are, however, heavily contaminated with RFI making it hard to recognise this characteristic shape. These sub-bands were therefore piped through {\tt pulse\_extract.py}, a code developed by \citet{mhl+18}, to group single pulse candidates and mitigate RFI, for instance by removing candidates detected at the same time with a similar S/N in several DM trials (i.e. narrow band RFI). The code reduced the number of candidates significantly but did not find any bursts from \frb. Finally, since we have an estimate of the arrival times of the bursts in the LOFAR band by extrapolating the arrival times of the Effelsberg bursts to LOFAR frequencies, a detailed inspection of the candidates at these times could be performed. Again, no bursts were found leading us to report a non-detection of bursts and burst counterparts of \frb\ with LOFAR.

\subsection{Simulations}\label{sec:sim}
Pulse profiles with large DMs are subject to a number of instrumental and propagation effects that are hard to disentangle at LOFAR frequencies. To quantify their influence on the sensitivity of the applied sub-band search, or in other words to investigate how well the search was able to recover initial burst energies, injection tests with fake single pulses in the LOFAR beamformed data were performed.

To this purpose, a Gaussian profile was injected into a copy of the beamformed data simultaneous with burst IX. The profile, with a width of 4\,ms, was scattered with a frequency-dependent pulse-broadening timescale determined for a DM of 188\,pc\,cm$^{-3}$ (100\,ms at 150\,MHz) while conserving the area of the profile. Then it was smeared and dispersed using the DM of \frb, before being piped through the sub-band search pipeline. The results were analysed in a similar fashion as described at the end of Sect. \ref{sec:analysis-lof} to see if the initial integrated flux of the pulse could be reconstructed. This process was repeated for an integrated flux ranging from 4 $\pm$ 2 to 16 $\pm$ 8\,Jy in steps of 0.4\,Jy when measured over a single sub-band. For the injection, a flat spectral index was assumed across the LOFAR band.

The lower flux density value used is the theoretical sensitivity limit of LOFAR at 150\,MHz calculated using Eq. 5 from \citet{kvh+16} with the same parameter dependencies as used in that paper. The exceptions are the position of \frb\ for the directional-dependent parameters, the number of summed stations $N_s$ being 46 (2 HBA fields $\times$ 23 core stations), a bad HBA tile fraction $\xi$ of 0.08, an RFI fraction $\zeta$ of 0.1, and bandwidth of 9.7\, MHz used in the sub-band search. This yields a sensitivity limit of 4 $\pm$ 2\,Jy at 150\,MHz to a 4\,ms burst at a 8-$\sigma$ confidence level. The above integrated flux errors arise from empirical results obtained by \citet{kvh+16}.

An intrinsic 4\,ms wide burst was injected, because the median of the widths of the Effelsberg bursts is 3.7\,ms. Assuming that the intrinsic widths of the bursts do not scale with frequency, LOFAR bursts of \frb\ will have an intrinsic width of $\sim$4\,ms. As long as the true intrinsic widths of LOFAR bursts are small ($<$ 10\,ms), such that the detected burst widths are dominated by scatter broadening, the results of the simulations remain essentially unchanged. Only if the intrinsic widths at 150\,MHz are of the same order as the scatter broadening will the results presented below change in an unfavourable way and LOFAR be less sensitive for scattered bursts.

Injected bursts were robustly detected at 150\,MHz from an initial integrated flux density of 11 $\pm$ 5\,Jy. Propagation effects, like scattering, thus reduced the sensitivity of the search pipeline by a factor of approximately three, though not evenly across the band. In the two sub-bands spanning the frequencies from 178 to 159\,MHz, the injected bursts were already detected at 10 $\pm$ 5\,Jy, but were not detected in the three lowest frequency sub-bands. This emphasises the importance of the use of the sub-band search technique; searches over the entire band might miss a pulse due to the large difference in sensitivity and effect of scattering across the LOFAR band. Since the theoretical sensitivity of LOFAR at 150\,MHz closely resembles the averaged theoretical sensitivity over the band, all forthcoming calculations are performed for a frequency of 150\,MHz. As the sensitivity for LOFAR, we therefore used the acquired value for the sensitivity limit of the sub-band search pipeline to scattered pulses of 11 $\pm$ 5\,Jy at 150\,MHz.

\section{Results}\label{sec:results}
All numbers are hereafter reported with 1-$\sigma$ error values if not mentioned otherwise.

\subsection{L-band bursts}
\begin{figure}
\resizebox{\hsize}{!}{\includegraphics{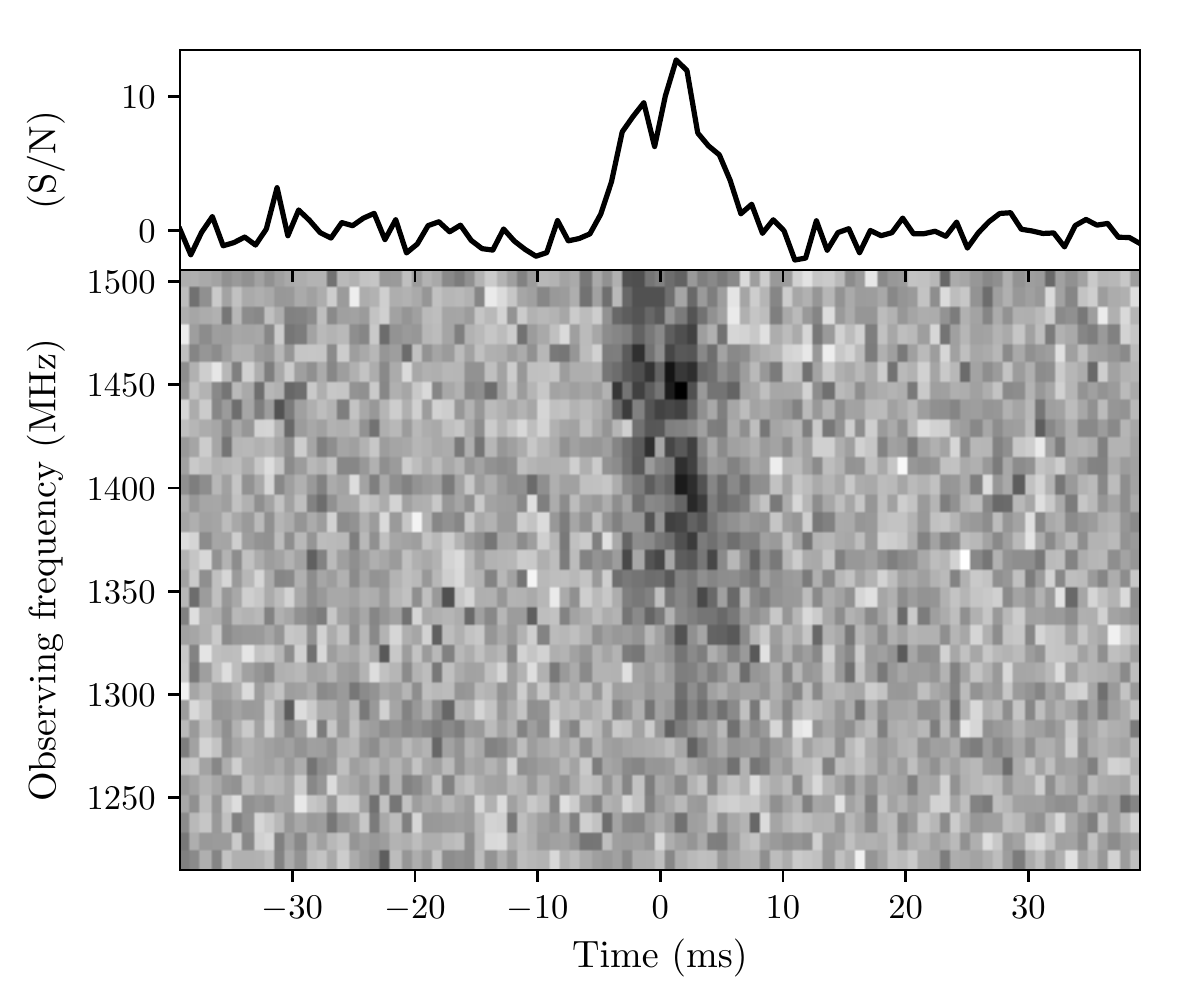}}
\vspace{-20pt}
\caption{Pulse profile and dynamic spectrum of burst VII in the PFFTS data showing its double peaked structure. The burst was dedispered to 560\,pc\,cm$^{-3}$, downsampled to a time resolution of 873.81\,$\mu$s, and centred at the arrival time of the burst in the PSRIX data given in Table \ref{tab:burstprop}.}
\label{fig:burst7}
\end{figure}

In total nine bursts were found with DMs consistent with \frb\ within the Effelsberg data. Table \ref{tab:burstprop} summarises their properties obtained from fitting Gaussian profiles to the data as described in Sect. \ref{sec:analysis}. Their fluences and pulse widths agree with the burst properties published in earlier work~\citep{sch+14,ssh+16-1,ssh+16-2,sbh+17,hds+17,lab+17,msh+18}, even though the widths are subject to significant intra-channel DM smearing of $\sim$1\,ms. From the published bursts it is apparent that pulse widths vary between 1 and 9\,ms and that the wider bursts often show a more complex structure~\citep{ssh+16-2,sbh+17,hss+18}. We also observed at least one burst with a more complex structure. Burst VII has a high enough S/N to partially resolve its burst profile and to show a double peak (Fig. \ref{fig:burst7}). The peaks are 3.6 $\pm$ 0.2\,ms apart and are most apparent for a dedispered profile with a DM of 560\,cm$^{-3}$, which is consistent with \citet{hss+18} reporting a DM of 560.57 $\pm$ 0.07\,pc\,cm$^{-3}$ to maximise \frb's burst structure.

\begin{table*}
\caption{Effelsberg L-band burst properties.}
\label{tab:burstprop}
\centering
\begin{tabular}{c c c c c c}
   \hline\hline
   \noalign{\smallskip}
   No. & TOA (MJD) & Integrated peak flux (Jy) & Fluence (Jy\,ms) & Gaussian FWHM (ms) & Integrated S/N\\
   \noalign{\smallskip}
   \hline
   \noalign{\smallskip}
   I    & 57635.267401552 & 0.12 $\pm$ 0.02 & 0.5 $\pm$ 0.1 & 4.6 $\pm$ 0.6 & 10\\
   II   & 57635.275278708 & 0.41 $\pm$ 0.06 & 0.7 $\pm$ 0.1 & 1.8 $\pm$ 0.1 & 22\\
   III  & 57635.297930512 & 0.14 $\pm$ 0.02 & 1.0 $\pm$ 0.2 & 7.1 $\pm$ 0.6 & 15\\
   IV   & 57635.319982700 & 0.17 $\pm$ 0.03 & 0.6 $\pm$ 0.1 & 3.7 $\pm$ 0.4 & 13\\
   V    & 57635.392856006 & 0.8  $\pm$ 0.1  & 2.4 $\pm$ 0.4 & 3.15 $\pm$ 0.08 & 54\\
   VI   & 57650.268119170 & 0.22 $\pm$ 0.03 & 0.7 $\pm$ 0.1 & 3.3 $\pm$ 0.3 & 16\\
   VII  & 57650.294844960 & 0.26 $\pm$ 0.04 & 1.9 $\pm$ 0.3 & 7.5 $\pm$ 0.3 & 28\\
   VIII & 57650.345421547 & 0.19 $\pm$ 0.03 & 0.9 $\pm$ 0.2 & 4.7 $\pm$ 0.4 & 17\\
   IX   & 57650.355313519 & 0.8  $\pm$ 0.1  & 2.7 $\pm$ 0.4 & 3.32 $\pm$ 0.07 & 61\\
   \noalign{\smallskip}
   \hline
\end{tabular}
\tablefoot{All reported errors are the one standard deviation errors and the given TOAs are barycentred and referenced to infinite frequency.}
\end{table*}

\subsection{Repeat rates}\label{sec:repeat_rates}
Effelsberg detected the nine bursts presented here from \frb\ in September 2016, whereas none were detected two months later. The overall averaged burst repeat rate for Effelsberg at 1.4\,GHz is therefore $R = 0.46 \pm 0.15$\,hr$^{-1}$ for bursts with an S/N above 10. As can be seen from Table \ref{tab:obsprop}, five and four bursts were detected on September 4 and 19 respectively, while no bursts were detected on November 13 and 14. This yields two repeat rates for September of $R_{\mathrm{Sep04}} = 1.1 \pm 0.5$\,hr$^{-1}$ and $R_{\mathrm{Sep19}} = 1.0 \pm 0.5$\,hr$^{-1}$. Having adopted a Poisson distribution for the burst detection probability to calculate the given errors~\citep{cp18}, we note that these rates are statistically consistent with each other. The same holds true for the November observations with no detections. We therefore argue a single rate to exist that is valid for the pair of observations in September and a second rate to exist for the pair of observations in November. The rates per month can then be combined to slightly reduce the uncertainties on the rates. This gives for the 8.3\,hr of September data $R = 1.1 \pm 0.4$\,hr$^{-1}$ compared to a repeat rate limit of $R < 0.3$\,hr$^{-1}$ (95\% confidence limit) for the non-detection in 11.1\,hr of November data.

In the same way can we derive repeat rate limits for LOFAR, where no bursts were detected in 8.9 and 9.4\,hr in September and November respectively. This results in a rate limit of $R < 0.3$\,hr$^{-1}$ (95\% confidence limit) for each period.

A similar variation in repeat rates as seen within the Effelsberg data was reported by \citet{lab+17}. The VLA detected nine bursts at 3\,GHz~\citep{clw+17} in September 2016, but did not find any during an earlier observing campaign in the same year. The bursts of \frb\ are thus correlated in time as was reported by \citet{ssh+16-2}, \citet{lab+17}, \citet{oyp18}, and \citet{shb+18}.

\subsection{Instantaneous spectral index limit}\label{sec:isi}
Despite extensive efforts, both a blind search for bursts and a targeted search for burst counterparts in the LOFAR data did not yield any bursts. Nevertheless, we can put constraints on their spectrum at 1.4\,GHz and 150\,MHz using this non-detection and the nine Effelsberg bursts. Therefore we consider not the bursts' fluxes but their fluences and assume their intrinsic widths not to scale with frequency. Since burst IX was the brightest detection, its fluence was taken as the reference fluence of Effelsberg at 1.4\,GHz. For the LOFAR fluence the sensitivity limit of the sub-band search was taken at 150\,MHz, determined in Sect. \ref{sec:sim}, multiplied with the adopted burst width of 4\,ms to obtain a fluence upper limit of 42\,Jy\,ms. Assuming the fluence $\mathcal{F}_{\nu}$ of bursts from \frb\ to scale with frequency as a power law ($\mathcal{F}_{\nu} \propto \nu^{\alpha}$), a limit on the spectral index $\alpha$ can be derived for its broad band instantaneous emission of $\alpha > -1.2 \pm 0.4$.
 
The above is the first simultaneous spectral index limit for bursts of \frb\ below 1\,GHz. For FRBs in general, \citet{kca+15} reported a spectral index limit of $\alpha > 0.1$ at 145\,MHz based on an FRB non-detection with a single international LOFAR station. Although this is a higher value, this limit is much more model-dependent. For instance, they used a constant volumetric rate of FRBs with redshift together with the assumption that FRBs are standard candles with broad band emission to derive their limit. Here we only assumed that \frb\ produces broad band bursts and absorbed the effects of propagation that could potentially explain a non-detection with LOFAR in the determination of its sensitivity limit. We also used the core stations of LOFAR in contrast to just one international station, gaining a factor of $\sim$14 in sensitivity and enabling us to detect fainter signals. Nevertheless, it is striking that both limits point towards a flat FRB emission spectrum below 1\,GHz, which could be an indication that the power law assumption for FRB emission does not hold true at low frequencies. This statement is backed-up by the recent results from \citet{sbm+18}. Using the Murchison Widefield Array (MWA) at $170-200$\,MHz, they did not find bursts simultaneously with bursts found by the Australian Square Kilometre Array Pathfinder (ASKAP) at L-band. They claim this to originate from a break in the intrinsic spectrum of FRBs at frequencies above 200\,MHz.

\section{Multi-frequency detection rate comparison}\label{sec:burst-comp}
In September 2016, \frb\ was observed to have an increased repeat rate where its bursts were observed at multiple frequencies and with several different telescopes~\citep[][this work]{sbh+17,clw+17}. Utilising the large bandwidth of the VLA, \citet{lab+17} showed that the individual burst spectra of \frb\ are poorly described by a single power law and better modelled with a Gaussian envelope of $\sim$500\,MHz and a varying central frequency. An intrinsic band-limited emission of the bursts would explain why no LOFAR bursts were detected simultaneously with the Effelsberg bursts. However, if the central frequencies of these envelopes shift into the LOFAR band, we may have expected to detect bursts non-simultaneously.

While the spectra of the individual bursts may be band-limited, the statistical distributions of burst centre frequencies and fluences may be frequency dependent, which would manifest observationally as a frequency-dependent occurrence rate of bursts of a given fluence. One possible mathematical description of this frequency dependence is a power law with a spectral index relating the normalisation of burst energy distributions at different frequencies. We refer to this as a statistical spectral index, which is distinct from an instantaneous spectral index. In practice a survey measures a burst detection rate above a sensitivity threshold. By measuring the detection rates from surveys conducted at multiple frequencies, one can measure this statistical spectrum, assuming observational biases such as propagation effects can be properly accounted for. Below is a simple mathematical derivation of the statistical spectral index. 

We assume that the statistical distribution of burst energies at a given radio frequency is described by a power law. Specifically we adopt the differential energy distribution from \citet{lab+17}: $dN/dE = T_{\rm obs}$\,$d\lambda/dE = AE^{\gamma}$, where $\lambda$ is the Poisson detection rate. The differential energy distribution has been recast as a differential rate distribution, because the key observable is number of bursts with a given energy per unit time. For a sensitivity-limited survey, most detections will be a factor of a few above the detection threshold. Directly comparing fluences between two burst samples from two surveys at different frequencies does not so much reflect the intrinsic source spectrum but rather the sensitivities of the surveys. Also, it is equally important to consider the total observing time required to detect a sample of bursts. If ten times as much time is needed at a higher frequency than at a lower frequency to observe a burst with similar fluence, it suggests that the bursts at the higher frequency are systematically weaker and a longer wait time is needed until a sufficiently bright burst occurs. 

Although burst energy is the more physically meaningful quantity, in practice we measure the fluence. Importantly, measuring a burst's energy requires being able to measure the bandwidth of the burst ($\Delta \nu$) and knowing the distance to the source ($D$) and the beaming factor of the emission ($\Omega$). Therefore, we will make a change of variables in the differential rate distribution from energy to fluence:
\begin{equation}
d\lambda = \left( \frac{\Omega}{4\pi} D^2 \Delta \nu \right)^{\gamma+1}d\mathcal{F}.
\end{equation}
We extend the formalism of \citet{lab+17} by assuming that the normalisation factors of the differential rate distributions ($A$) at two frequencies are related through a power law parameterised by a statistical spectral index ($\alpha_s$), that is, bursts with fluences $\mathcal{F}_{\nu,1}$ and $\mathcal{F}_{\nu,2}$ observed at frequencies $\nu_1$ and $\nu_2$ are related by $\mathcal{F}_{\nu,1}/\mathcal{F}_{\nu,2} = (\nu_1/\nu_2)^{\alpha_s}$. 
We enforce the shift in the normalisation by relating the fluences at $\nu_1$ and $\nu_2$ for the same rate ($d\lambda_{\nu1}/dN = d\lambda_{\nu2}/dN$) with the statistical spectral index. 

In practice we usually calculate the rate of all burst detections above a minimum detectable fluence. This can be obtained by integrating the differential rate distribution from the lowest detectable fluence $\mathcal{F}_{\rm min}$ to a maximum burst fluence $\mathcal{F}_{\rm max}$. To avoid the need for an absolute normalisation, we take the ratio between two observed burst rates at two different frequencies. This ratio can be further simplified when we assume that $\mathcal{F}_{\rm max} \gg  \mathcal{F}_{\rm min}$. We also take $\gamma < -1$, such that the detection rate $\lambda$ is dominated by burst fluences close to the fluence detection threshold $\mathcal{F}_{\rm min}$, to obtain
\begin{equation}\label{eq:ssi}
\frac{\lambda_1}{\lambda_2} \approx \left(\frac{\nu_1}{\nu_2}\right)^{-\alpha_s\gamma}\left(\frac{\mathcal{F}_{\nu\rm 1,min}}{\mathcal{F}_{\nu\rm 2,min}}\right)^{\gamma+1}.
\end{equation}
We assumed that the emission beaming factor $\Omega$ is frequency independent when deriving the above expression.
With Eq.~\ref{eq:ssi} a statistical spectral index can be obtained from the observed burst rates at two frequencies and the corresponding fluence complete limits of the observations.

Importantly, several assumptions are built into this derivation. First, we assume that the rates at two different frequencies are scaled solely by the statistical spectral index and not caused by an intrinsic change in the rate of bursts generated at those frequencies. Second, we assume that $\gamma$ is constant with frequency and time, which is consistent with \citet{lab+17}, who derived a value for $\gamma$ of $-1.7$. Finally, we assume that $\lambda$ is constant in time during September 2016. We know that this it not the case for \frb\ over long timescales. In this work we present a markedly different rate in observations separated by roughly two months. Nevertheless, we do know that the detection rate was higher than average during September 2016 from the near daily observation campaign with the VLA~\citep{clw+17}. Furthermore, most of those observation sessions were 120 minutes long within which zero or one burst was detected, suggesting that the detection rate did not vary significantly during those two weeks. One possible exception is the 54-minute session on 2 September 2016 during which two bursts were detected. 

The sample of previously published bursts detected in this period come from the VLA at 3\,GHz \citep[i.e. $2.5-3.5$\,GHz band,][]{lab+17}, the Green Bank Telescope (GBT) at 2\,GHz \citep[i.e. $1.6-2.4$\,GHz band,][]{sbh+17}, and we add to them the Effelsberg data at 1.4\,GHz (i.e. $1.2-1.5$\,GHz band), presented in this paper. This gives three burst detection rates for the three central frequencies of the telescopes above. The derived rates are $R_{\rm VLA} = 0.3 \pm 0.1$\,hr$^{-1}$, $R_{\rm GBT} = 0.5 \pm 0.3$\,hr$^{-1}$ , and $R_{\rm EFF} = 1.1 \pm 0.4$\,hr$^{-1}$. The Effelsberg and VLA rates are consistent at the $1.5-\sigma$ level, whereas the GBT rate is consistent with both the Effelsberg and VLA rates.

Applying Eq. \eqref{eq:ssi} to the measured detection rates from the VLA and GBT and using the Effelsberg bursts as a reference, we find two values for $\alpha_s$: $\alpha_{s,\rm\text{ VLA / EFF}} = -1.3_{-0.6}^{+0.5}$ and $\alpha_{s,\rm\text{ GBT / EFF}} = -2.5_{-1.5}^{+1.2}$. For our analysis we adopted the fitted value of $\gamma = -1.7$ taken from \citet{lab+17} together with their definition of the fluence completeness limit. This limit, being defined as an effective detection limit of 0.9 times the weakest burst detected, resulted in the following fluence completeness limits used for the second term on the right of Eq. \eqref{eq:ssi}: $\mathcal{F}_{\nu\rm, min}^{\rm VLA} = 0.2$\,Jy\,ms, $\mathcal{F}_{\nu\rm, min}^{\rm GBT} = 0.1$\,Jy\,ms, and $\mathcal{F}_{\nu\rm, min}^{\rm EFF} = 0.5$\,Jy\,ms, respectively. Although the errors on the statistical spectral indices are large, they tentatively suggest that \frb\ has a negative spectral index for its overall distribution of burst energies versus frequency. 

Now we calculate a limit for the statistical spectral index of potential bursts at 150\,MHz and those at 1.4\,GHz with the September repeat rate of LOFAR from Sect. \ref{sec:repeat_rates}. Adopting the fluence upper limit of LOFAR of 42\,Jy\,ms, we obtain $\alpha_{s,\rm\text{ LOF / EFF}} > -0.5_{-0.2}^{+0.2}$. This is a much flatter spectral index than at the approximately gigahertz frequencies and is marginally inconsistent with the previously mentioned values for $\alpha_{s,\rm\text{ VLA / EFF}}$ and $\alpha_{s,\rm\text{ GBT / EFF}}$. The statistical spectral indices for \frb\ thus seem to indicate a flattening of its spectrum at low frequencies, which is in line with the results presented in Sect. \ref{sec:isi}.

To better quantify this possible flattening of the spectra of bursts from \frb\ below 1\,GHz, more precise statistical spectral indices must be obtained. More LOFAR observations simultaneous with multiple higher frequency observations are therefore needed to further constrain the repeat rate of \frb\ at 150\,MHz and observing frequencies of the other observatories. Alternatively, the sensitivity of LOFAR can be increased to yield better constrained repeat rate limits and potentially a burst from \frb. This can be done by performing observations with not just LOFAR's core stations but with its remote stations as well, or by obtaining LOFAR data that can be coherently dedispersed to gain a wider usable bandwidth.

\section{Summary and conclusions}\label{sec:conclusion}
In late 2016 we performed roughly 20 hours of simultaneous observations on \frb\ with Effelsberg and LOFAR. The observation time was divided over four sessions. During two sessions in September, nine bursts from \frb\ were detected at 1.4\,GHz, while no bursts were detected in two sessions in November. With a newly designed single pulse sub-band search method, no bursts were detected at 150\,MHz in the LOFAR beamformed data. From this we derived a spectral index limit for the fluence of the    instantaneous broad band emission of \frb\ between 1.4\,GHz and 150\,MHz of
\begin{equation*}
\alpha > -1.2 \pm 0.4.
\end{equation*}

In order to obtain this instantaneous spectral index, simulations were performed to determine the response of the sub-band search method to strongly scattered pulses. We showed that its sensitivity is a factor of three lower than the theoretical sensitivity of LOFAR due to the intensity drop of the bursts from \frb\ when scattered to a width of 100\,ms at 150\,MHz.

Despite extensive efforts to detect a burst from \frb\ at multiple frequencies~\citep{ssh+16-2,sbh+17,hds+17}, this has only been achieved three times~\citep{sbh+17,lab+17}, which suggests that simultaneous detections at frequencies separated by 1\,GHz are unlikely. Nevertheless, we could have found non-simultaneous bursts, the lack of which led us to calculate a statistical spectral index to estimate the probability of detecting bursts given the rate of Effelsberg bursts at 1.4\,GHz. This was done by obtaining statistical spectral indices for bursts detected at different frequencies at the VLA, GBT, and Effelsberg using the repeat rate ratio analysis presented here.

Three statistical spectral indices were calculated:
\begin{align*}
\alpha_{s,\rm\text{ VLA / EFF}} &= -1.3_{-0.6}^{+0.5},\\
\alpha_{s,\rm\text{ GBT / EFF}} &= -2.5_{-1.5}^{+1.2},\\
\alpha_{s,\rm\text{ LOF / EFF}} &> -0.5_{-0.2}^{+0.2}.
\end{align*}
The indices calculated at frequencies above 1\,GHz are consistent within the uncertainties, while below 1\,GHz a flattening in the statistical spectral index may occur. Whether this is a real feature of the distribution is yet unclear. More observations, preferably more sensitive, coherently dedispered LOFAR observations, are needed to provide clarity about the exact value and any dependence on frequency of the spectral index of \frb. This is achieved by applying the repeat rate ratio analysis on observations performed simultaneously over a wide frequency range and long time span.

~~~~~

Very recently, a second repeating FRB has been discovered by \citet{CHIME}. It will be interesting to repeat the experiment for this source and to compare the results with the conclusions reached here.


\begin{acknowledgements}
This paper is based on observations with the 100-m telescope of the MPIfR (Max-Planck-Institut f\"ur Radioastronomie) at Effelsberg and (in part) on data obtained with the International LOFAR Telescope (ILT) under project code LC6\_009. LOFAR~\citep{hwg+13} is the Low Frequency Array designed and constructed by ASTRON. It has observing, data processing, and data storage facilities in several countries, which are owned by various parties (each with their own funding sources), and that are collectively operated by the ILT foundation under a joint scientific policy. This work is part of the research programme TOP1EW.14.106 with project number 614.001.454, which is (partly) financed by the Netherlands Organisation for Scientific Research (NWO). L.G.S. acknowledges financial support from the ERC Starting Grant BEACON under contract number 279702, as well as the Max Planck Society.
\end{acknowledgements}


\vspace{1cm}
\bibliographystyle{aa} 
\bibliography{../masterbib} 

\end{document}